\documentclass[aps,prb,showpacs,superscriptaddress,floatfix,twocolumn]{revtex4}
\usepackage{amsmath}
\usepackage{citesort}
\usepackage{graphicx}
\begin{document}
\title{ARPES and optical renormalizations: phonons
  or spin fluctuations}
\author{E. Schachinger}
\email{schachinger@itp.tu-graz.ac.at}
\homepage{www.itp.tu-graz.ac.at/~ewald}
\affiliation{Institut f\"ur Theoretische Physik, Technische Universit\"at
Graz\\A-8010 Graz, Austria}
\author{J.J. Tu}
\affiliation{Department of Physics and Optical Science, University of
North Carolina at Charlotte, Charlotte, North Carolina 28223}
\author{J.P. Carbotte}
\affiliation{Department of Physics and Astronomy, McMaster University,\\
Hamilton, Ontario, Canada L8S 4M1}
\date{\today}
\begin{abstract}
  Improved resolution in both, energy and momentum in ARPES-data has
  lead to the establishment of a definite energy scale in the
  dressed quasiparticle dispersion relations.
  The observed structure around $80\,$meV has been taken as evidence
  for coupling to phonons
  and has re-focused the debate about the mechanism of superconductivity
  in the cuprates. Here we address the relative merits of
  phonon as opposed to spin fluctuation mechanisms. Both
  possibilities are consistent with ARPES. On the other hand,
  when the considerations are extended to infrared optical data, a
  spin fluctuation mechanism provides a more natural
  interpretation of the combined sets of data in
  Bi$_2$Sr$_2$Ca Cu$_2$O$_{8+\delta}$ (Bi2212).
\end{abstract}
\maketitle
\section{Introduction}

With the advent of improved resolution, both in energy and momentum,
several angular resolved photo emission spectroscopy (ARPES)
groups have measured the renormalized quasiparticle
energies and/or lifetimes\cite{kaminski,valla,johnson,kaminski1,%
  lanzara,bogdanov,valla1,valla2}
in the cuprates. The data shows that a definite energy
scale exists for the renormalization. This energy scale has tentatively
been assigned by some\cite{lanzara,bogdanov}
to interactions of the charge carriers with phonons.
This provocative possibility has again brought to the forefront
the open question of mechanism in the cuprates and in particular
the possibility that the electron-phonon interaction plays
an important role. While the electron-phonon interaction is
widely believed to cause superconductivity in conventional
materials, the gap symmetry in the cuprates is
$d$-wave\cite{mars6,hardy,shen,wollm,tsuei}
rather than $s$-wave
which means that it is the projection
of the electron phonon interaction on the $d$-channel which
enters the equation for the critical temperature $T_c$.
This would imply that some of the details of the electron-phonon
interaction would be drastically different in the oxides
compared to
conventional metals. While conventional superconductors do
exhibit anisotropy\cite{tomlinson,carb2}
in their superconducting gap as a function of momentum {\bf k}
on the Fermi surface, the $s$-channel always overcomes the $d$-channel
interaction. Also in dirty $s$-wave materials with the mean free
path $\ell$ smaller than the coherence length $\xi_0$, the anisotropy
is washed out and the gap becomes isotropic. On the other hand,
in the pure limit there are many ways
in which this gap anisotropy manifests itself in a conventional
superconductor.\cite{carb2}
One example is the low temperature $(T)$ electronic
specific heat which exhibits the expected exponential dependence
$\exp\{-\Delta/T\}$ only for temperatures much lower than the
lowest gap $(\Delta)$ in the system.

Arguments for why the form of the electron-phonon interaction can
be very different in the oxides as compared with conventional
metals have been presented in the literature. Some of the critical
ideas have been reviewed by Kuli\'c.\cite{no15}
The main argument for our
purpose here is that the electron-phonon interaction becomes
strongly peaked in the forward direction when corrections for
strong correlations are properly accounted for through charge
vertices.\cite{zeyher,kulic} Another related idea is that
because of the large static dielectric function in the oxides the
screening is greatly reduced as compared with ordinary metals.
This also leads to enhanced forward scattering as can be
seen\cite{weger,weger1} from the following simple argument.
The Fourier transform of the bare Coulomb potential
diverges like $1/q^2$ for small {\bf q}, here {\bf q} is the momentum
transfer.
Screening eliminates this singularity and the screened potential goes,
instead, like $1/(q^2+\lambda^2_{\rm FT})$, where $\lambda_{\rm FT}$
is an inverse screening length. For small $\lambda_{\rm FT}$ the
increased forward
scattering has the general tendency of increasing the
weighting of the higher harmonics
in its expansion. For example the expansion of a delta function
$\delta({\bf q})$ has equal weight in each spherical harmonics.%
\cite{kulic1} Strong forward scattering can in fact lead to
$d$-wave superconductivity which is not suppressed by the
strong Coulomb repulsion in contrast to the $s$-wave channel.%
\cite{kulic1,kulic2}

While the possibility of phonon induced $d$-wave superconductivity
cannot be eliminated on general grounds, it has not been favored
in much of the literature. While there is no consensus on mechanism, many
workers believe instead in an electronically driven scenario: for instance,
the $t-J$ model can exhibit superconductivity.\cite{sorella}
A spin fluctuation
mechanism, such as is envisaged in the Nearly Antiferromagnetic
Fermi Liquid (NAFFL) model of Pines and coworkers\cite{pines1,pines2}
is also electronic in nature with the exchange of spin fluctuations
rather than phonons. Another electronic mechanism is the
Marginal Fermi Liquid (MFL) model\cite{varma1,varma2,varma3}
which in its original form had $s$-wave symmetry.

In this paper we attempt to understand the renormalization effect
observed in ARPES within a boson exchange mechanism. We will
consider explicitly both, phonon and spin fluctuation exchange.
We will also try to reconcile within the same model, both, ARPES
data on equilibrium quasiparticle properties and the optical conductivity.
Infrared measurements in the cuprates have produced a wealth
of information on charge dynamics in the oxides.\cite{puchkov,%
  timusk} They have played a
key role in our present understanding of the microscopic
nature of the cuprates. In particular, $c$-axis infrared conductivity
data have given detailed spectroscopic information on the
pseudogap\cite{timusk} which has been widely viewed as a direct manifestation
of strong correlation effects. Recently, improvements in resolution
have also been achieved in
  $ab$-plane measurements.\cite{tu}
  
From reflectance
measurements as a function of frequency it is possible to extract
separately
the real and the imaginary part of the conductivity $\sigma(T,\omega)$
as a function of $\omega$ for a fixed temperature $T$. Within a
generalized Drude model, in which the optical scattering rate
and optical effective mass acquire a frequency dependence, we can define
an isotropic optical scattering time $\tau_{op}(T,\omega)$ as
$\tau^{-1}_{op} = (\Omega^2_p/4\pi)\Re{\rm e}\sigma^{-1}(T,\omega)$,
where $\Omega_p$ is the plasma frequency.
This quantity can be determined
from the optical sum rule on the real part of $\sigma(T,\omega)$,
namely $\int_0^\infty d\omega\sigma_1(T,\omega) = (\Omega^2_p/8)$.
The optical scattering rate defined above is related to the
quasiparticle lifetime which may be considered to be
more fundamental
as they directly define quasiparticle motion.
While these two lifetimes are related, they are by no means identical,
however, except for a simplified case of isotropic elastic impurity
scattering. Even in this case, if the electronic density of states
has an important dependence on energy or the elastic scattering is
anisotropic, quasiparticle and transport scattering rates are no
longer the same. Note that, in as much as the in-plane conductivity
is isotropic, the optical scattering
rate is also isotropic and represents a weighted average of the more
fundamental momentum dependent quasiparticle scattering. While
there have been
some attempts in the literature\cite{kaminski}
to compare both sets of data, i.e.:
ARPES and optical directly, we will emphasize here that,
because they are not so
simply related, they cannot easily be compared.
On the other hand, any viable
microscopic model needs to be able to provide a unifying
description of both sets of experiments. In this regard, we will
emphasize that in addition to the differences between quasiparticle
and transport scattering rate which we have already discussed, they
differ in another fundamental way. In a boson exchange mechanism
the charge carrier-exchange boson spectral density (denoted by
$\alpha^2F(\omega)$ in the explicit case of phonons) which describes
the equilibrium properties can be quite different from the
corresponding transport spectral density, usually denoted by
$\alpha^2_{tr}F(\omega)$.\cite{tomlinson1,leung}
The origin of this fundamental difference
lies in a well-known factor
$(1-\cos\theta)$, where $\theta$ is the angle between initial and
final electron momentum undergoing a scattering process. For
quasiparticle properties such as its lifetime, only the
probability that an electron of momentum {\bf k} leaves the state
$\vert{\bf k}\rangle$ is relevant, while for transport, for instance
for the d.c. resistivity, it is also important where it ends up
since backward scattering depletes the current more than
forward scattering. This basic physics is accounted for by the
$(1-\cos\theta)$ factor which eliminates forward scattering
$(\theta = 0)$ and weights with a factor two backward scattering
$(\theta = \pi)$.
Serious estimates\cite{no15,zeyher,kulic}
of these two spectral densities
for a model electron-phonon interaction in the poor screening
regime with explicit inclusion of correlation effects, have lead
to an estimate that $\alpha^2_{tr}F(\omega)$ may be smaller
by about a
factor of $1/3$ compared with $\alpha^2F(\omega)$, the corresponding
quasiparticle quantity. On the other hand, in the NAFFL model
the interaction with spin fluctuations is believed to
strongly peak around
$(\pi,\pi)$ in momentum. This means that the scattering
potential itself weights more strongly the
backward scattering processes, and so
we expect that the transport spectral density
will be larger than its quasiparticle counterpart.
The above arguments immediately
suggest that phonons effects could very well show up
prominently in quasiparticle properties while at the same time
play only a minor role in transport, i.e.: in the infrared
conductivity and vice versa for spin fluctuations.

In section II we make general remarks about the renormalization
effects due to boson exchange mechanisms based on perturbation theory. In
particular we contrast the electron-phonon case with spin
fluctuations. In section III we deal with fits to the ARPES data
in Bi2212 and in section IV we consider, in addition, optical
conductivity data. We describe the constraints on microscopic
models that a fit to both sets of data imposes in the specific case of
Bi2212. In section V we draw conclusions.

\section{General results}

For an electron-phonon system, the renormalization of the electronic
quasiparticles follows from a knowledge of the electron-phonon
spectral density $\alpha^2F(\omega)$. This function which depends
only on frequency, and which is limited in range to the
maximum phonon frequency, contains all of the complicated information
about electronic band wave functions, dispersion relation, phonon
dynamics, and electron-lattice vibration coupling
which is needed to compute
self energy effects. In lowest order perturbation theory, the
quasiparticle scattering rate $\tau^{-1}_{qp}(\omega)$ at energy
$\omega$ is given by\cite{allen}
\begin{equation}
  \label{eq:1}
  \tau^{-1}_{qp}(\omega) = 2\pi\int\limits_0^\omega\!d\nu\,
  \alpha^2F(\nu).
\end{equation}
For a $\delta$-function Einstein phonon at $\omega_E$ of the form
$\alpha^2F(\omega) = A\delta(\omega-\omega_E)$
\begin{equation}
  \tau^{-1}_{qp}(\omega) = \left\{\begin{array}{rl}
        2\pi A & \omega > \omega_E\\
        0 & \omega < \omega_E,
        \end{array}\right .
  \label{eq:2}
\end{equation}
and we see that $\omega_E$ sets the energy at which $\tau^{-1}_{qp}%
(\omega)$ jumps from zero to a finite value. For an $\alpha^2F(\omega)$
distributed in energy, the rise will be more gradual but for
$\omega\ge\omega_D$, with the Debye energy $\omega_D$,
the scattering rate will, again, become constant.

The expression for the optical scattering rate, on the other hand,
which is also obtained in lowest
order perturbation theory is\cite{allen}
\begin{equation}
  \label{eq:3}
  \tau^{-1}_{op}(\omega) = {2\pi\over\omega}\int\limits_0^\omega\!d\nu\,
  (\omega-\nu)\alpha^2_{tr}F(\nu),
\end{equation}
which gives $\tau^{-1}_{op}(\omega) = 0$ for
$\omega\le\omega_E$ in an Einstein model, but now for
$\omega > \omega_E$
\begin{equation}
  \label{eq:4}
  \tau^{-1}_{op}(\omega) = 2\pi\left(1-\frac{\omega_E}{\omega}
    \right) A_{tr},
\end{equation}
which starts at zero with $\omega=\omega_E$ and gradually increases
towards $2\pi A_{tr}$ with increasing $\omega$. This is quite
distinct from the abrupt jump to $2\pi A$ at $\omega=\omega_E$
found for the quasiparticle scattering rate. Here the subscript
`$tr$' denotes transport as opposed to equilibrium properties.
It is clear
that the singularity in $\tau^{-1}(\omega)$
 associated with the boson energy scale $\omega_E$
is more significant in the quasiparticle (equilibrium case) than
in the transport scattering rate.
There is yet another difference between quasiparticle and
transport rates. If we denote the electron-phonon
interaction matrix element by $g_{{\bf k},{\bf k}',\nu}$ for
electrons scattering from {\bf k} to ${\bf k}'$ with a phonon
of energy $\omega_\nu({\bf k}-{\bf k}')$ ($\nu$ is a branch
index) the quasiparticle spectral density is\cite{grimvall,carb1,%
  mars6}
\begin{equation}
  \label{eq:5}
  \alpha^2F(\omega) = \sum\limits_\nu\left\langle\left\langle
      \left\vert g_{{\bf k},{\bf k}',\nu}\right\vert^2
      \delta(\omega-\omega_{\nu}({\bf k}-{\bf k}'))
      \right\rangle\right\rangle,  
\end{equation}
with the double average over the Fermi surface denoted by
$ \left\langle\left\langle\cdots\right\rangle\right\rangle$
\begin{equation}
  \label{eq:6}
 \left\langle\left\langle G \right\rangle\right\rangle
 = \frac{\sum\limits_{{\bf k},{\bf k}'}G_{{\bf k},{\bf k}'}
   \delta(\varepsilon_{\bf k}-\mu)\delta(\varepsilon_{{\bf k}'}-\mu)
   }{\sum\limits_{\bf k}\delta(\varepsilon_{\bf k}-\mu)},
\end{equation}
where $\mu$ is the chemical potential, so all integrals are pinned
at the Fermi surface. The optical spectral weight is
related\cite{tomlinson,leung,grimvall} instead to
\begin{equation}
  \label{eq:7}
  \alpha^2_{tr}F(\omega) \propto \sum\limits_\nu\left\langle\left\langle
      \left\vert g_{{\bf k},{\bf k}',\nu}\right\vert^2
      \left(v_{\bf k}-v_{{\bf k}'}\right)^2
      \delta(\omega-\omega_{\nu}({\bf k}-{\bf k}'))
      \right\rangle\right\rangle.
\end{equation}
The extra factor of $v_{\bf k}-v_{{\bf k}'}$ with $v_{\bf k}$ the
velocity of the electron {\bf k} weights forward scattering by zero
and emphasizes most backward collisions.

The functions $\alpha^2F(\omega)$ of Eq.~(\ref{eq:5}) and
$\alpha^2_{tr}F(\omega)$ of Eq.~(\ref{eq:7}) are isotropic,
momentum averages, given by the right hand side of
each of these equations. Both involve a double integral over
initial and final states $\vert{\bf k}\rangle$ and $\vert{\bf k}'\rangle$.
The full complicated momentum dependence of the electron-%
phonon coupling $g_{{\bf k},{\bf k}',\nu}$ and of the
phonon dispersion $\omega_\nu({\bf k}-{\bf k}')$ is included
with no approximations, although after the double summation
over momentum, indicated in Eq.~(\ref{eq:6}),
$\alpha^2F(\omega)$ is now isotropic, dependent only on frequency.
This is the function that
enters isotropic Eliashberg theory. An
anisotropic formulation of the Eliashberg equations\cite{daams}
also exists and in this case a directional electron-phonon spectral
density distinct for each electron state $\vert{\bf k}\rangle$
on the Fermi surface, which we denote by $\alpha^2_{\bf k}F(\omega)$,
would replace its Fermi surface average:
\begin{equation}
  \alpha^2F(\omega) = \frac{\sum\limits_{\bf k}\alpha^2_{\bf k}F(\omega)}
  {\sum\limits_{\bf k}\delta(\varepsilon_{\bf k}-\mu)}.
  \label{eq:aniso}
\end{equation}
To calculate $\alpha^2F(\omega)$ from first principles is difficult and
complex.\cite{carb1} It can, fortunately, be measured directly from tunneling
data through consideration of current voltage characteristics.%
\cite{mcmillan,mcmillan1}
The resulting function of $\omega$ often, but
not always, looks qualitatively very much like the phonon
frequency distribution
$F(\omega)$ but this does not imply that momentum dependence
of coupling and/or of phonons is not included in Eq.~(\ref{eq:5}).
Any nesting effects such as are present in the nested Fermi
liquid theory of Virosztek and Ruvalds\cite{ru1,ru2}  and in the work of
Savrasov and Andersen\cite{savr} are fully incorporated on the right
hand side of Eq.~(\ref{eq:5}). It is only because we take the
left hand side from experiment that these important issues do
not become prominent in our work.

A further complication arises in the case of coupling to spin
fluctuations.\cite{branch} In this case the magnetic spin susceptibility
$\chi_{{\bf k},{\bf k}'}(\omega)$ is involved as well as its coupling
to charge. The susceptibility in the cuprates is sharply peaked in
momentum space at the antiferromagnetic wave vector
${\bf k}-{\bf k}'$ (momentum transfer) for electron scattering
from $\vert{\bf k}\rangle$ to $\vert{\bf k}'\rangle$. In
this case, what enters Eliashberg theory\cite{branch} is the complex
function $\chi_{{\bf k},{\bf k}'}(\omega)$ weighted by electron
coupling. For instance, in the NAFFL model\cite{pines1}
\begin{equation}
  \label{eq:10b}
  \chi_{{\bf k},{\bf k}'}(\omega) = 
  {\chi_{\bf Q}\over 1+\xi^2({\bf q}-{\bf Q})^2-i{\omega\over%
  \omega_{MMP}}},\quad q_x,\, q_y > 0,
\end{equation}
with ${\bf q} = {\bf k}-{\bf k}'$,
$\chi_{\bf Q}$ the static susceptibility, $\xi$ the
magnetic coherence length, ${\bf Q}=(\pi,\pi)$ the commensurate
antiferromagnetic momentum, and $\omega_{MMP}$ the characteristic
energy of the spin fluctuations at the point {\bf Q}.
A Fermi surface to Fermi surface approximation does
not naturally reveal itself since this may not include
transitions with ${\bf k}-{\bf k}' = (\pi,\pi)$ for which
the susceptibility is the largest. However, one can still
introduce, as an approximation, some average effective
electron-spin fluctuation spectral function denoted by
$I^2\chi(\omega)$ which can be used in the
isotropic version of the Eliashberg equations.
It represents the appropriate average of the susceptibility that
enters superconductivity. Its value is to be determined from
consideration of experimental data.
As described by Carbotte {\it et al.}\cite{schach4} this
isotropic spectral density is given by
the second derivative of $\omega$ times the optical scattering rate
$\tau_{op}^{-1}(\omega)$. Recall,
$\tau_{op}^{-1}(\omega) = (\Omega_p^2/4\pi)\Re{\rm e}\sigma^{-1}(\omega)$,
where $\sigma(\omega)$ is the in-plane optical conductivity. This
conductivity is isotropic in tetragonal systems. For the
orthorhombic case, $\sigma(\omega)$ with ${\bf E}\parallel a$ and
${\bf E}\parallel b$ can be different. Defining
\begin{equation}
  \label{eq:10a}
  W(\omega) = {1\over 2\pi}{d^2\over d\omega^2}
  \left(\frac{\omega}{\tau_{op}(\omega)}\right),
\end{equation}
the approximate relation $I^2_{tr}\chi(\omega)\simeq W(\omega)$
holds.\cite{mars4} We emphasize that in terms of microscopic
theory $W(\omega)$ is
related to an appropriate average of the full momentum and
energy spin susceptibility $\chi_{{\bf k},{\bf k}'}(\omega)$
(Eq.~\eqref{eq:10b}).
Recently, infrared optical data
has been used very effectively to obtain $I^2_{tr}\chi(\omega)$ in
the oxides and we will use these in our work.\cite{schach4,schach7,%
schach5,schach8}
In this way we
circumvent all of the complications that would arise in a first
principle calculation of the average effective function
$I^2_{tr}\chi(\omega)$. We will return to this important point later in
equation (\ref{eq:10}).
We  note in passing that $W(\omega)$ very often
exhibits a pronounced resonance like structure particularly
if $\tau^{-1}_{op}(\omega)$ data at low temperatures in the
superconducting state is analyzed. In the following we use
the expression `optical resonance' in reference to such a structure.
In most materials analyzed so far, this optical resonance has a
spin resonance equivalent measured using inelastic neutron
scattering. We will discuss this in more detail in Sec.~\ref{sec:4}.

In some sense the opposite extreme to
the NAFFL model of Pines and coworkers\cite{pines1,pines2}
in which the interaction
peaks at $(\pi,\pi)$ is the MFL model of
Varma {\it et al.}\cite{varma1,varma2,varma3}
in which the interaction with the fluctuation
spectrum is thought of as momentum independent in a first
approximation. As originally conceived, this model leads to
an $s$-wave superconducting gap rather than $d$-wave as is now generally
believed to be the case. This represents a real limitation for the
model. Nevertheless, it has been very useful in correlating much data
on the normal state transport properties in the cuprates.

In as much as it is only the
magnitude and energy dependence of the resulting average spectral
function $\alpha^2F(\omega)$ ($I^2\chi(\omega)$)
that matter, these may not be very different between
the MFL and NAFFL models. Both have an average spectral density which
is reasonably flat as a function of energy and which extends to some
high frequency of order several hundred meV. Of course, if
anisotropies on the Fermi surface were to be considered, the
directional $\alpha^2_{\bf k}F(\omega)$ ($I^2_{\bf k}\chi(\omega)$),
Eq.~\eqref{eq:aniso}, would be expected to vary strongly in the
NAFFL model and not
in the MFL model, but here, for simplicity, we have assigned to each
electron the same average interaction. For phonons in conventional
metals the mass renormalization parameter
$\lambda_{\bf k} = 2\int_0^\infty\!d\omega\,\alpha^2_{\bf k}F(\omega)/%
\omega$ can vary by several tens of percentage points over the Fermi
surface.\cite{carb2} For the NAFFL model based on the susceptibility
\eqref{eq:10b}, the variations are much larger,
and can be of the order of a factor of
two or three.\cite{branch} It is clear that ARPES data in a
particular direction could considerably under or overestimate
the average renormalizations. In this sense optics is better.

We should make one further remark. In
conventional materials for which $\alpha^2F(\omega)$ and
$\alpha^2_{tr}F(\omega)$ have been calculated from
band structure and phonon information usually obtained by inelastic
neutron scattering, they have been found to differ\cite{tomlinson,%
  carb2,tomlinson1,leung} in shape as
a function of $\omega$ and in size. But
these differences have often been overlooked in the literature.
In the work of Kuli\'c and Zeyher \cite{no15,zeyher,kulic}
the corresponding differences are large. These authors
considered directly the cuprates and try to account seriously for the
ionicity and the increased effect of correlations.
As described in the introduction, they find
that the electron-phonon interaction in this case is strongly
peaked in the forward direction and this leads to large differences
between $\alpha^2F(\omega)$ and $\alpha^2_{tr}F(\omega)$. They
estimate $\alpha^2F(\omega)$ to be about a factor of three larger
than $\alpha^2_{tr}F(\omega)$. The consequence of this is that
phonons would show up much more prominently in quasiparticle
properties (equilibrium) than in the corresponding transport property. In
particular, at large $\omega$, $\tau^{-1}_{qp}(\omega)$
and $\tau^{-1}_{op}(\omega)$ (Eqs.~(\ref{eq:2}) and (\ref{eq:4}),
respectively) would approach each other if $\alpha^2F(\omega)$ and
$\alpha^2_{tr}F(\omega)$ were identical, but in fact
$\tau^{-1}_{op}(\omega)$ is expected to be larger than
$\tau^{-1}_{qp}(\omega)$. The opposite would hold for the NAFFL
model which emphasizes backward rather than forward scattering
because the interaction peaks at $(\pi,\pi)$.
In this case spin fluctuations
should show up more prominently in transport than in
equilibrium properties. Even though these differences are hard
to quantify, the general qualitative features of
transport as compared with quasiparticle spectral weight
just described must be kept in mind when analyzing data.
In the MFL model we expect smaller differences between
quasiparticle and transport spectral densities because the
underlying interaction is assumed to be approximately momentum
independent.

\begin{figure}
  \vspace*{-4mm}
  \includegraphics[width=9cm]{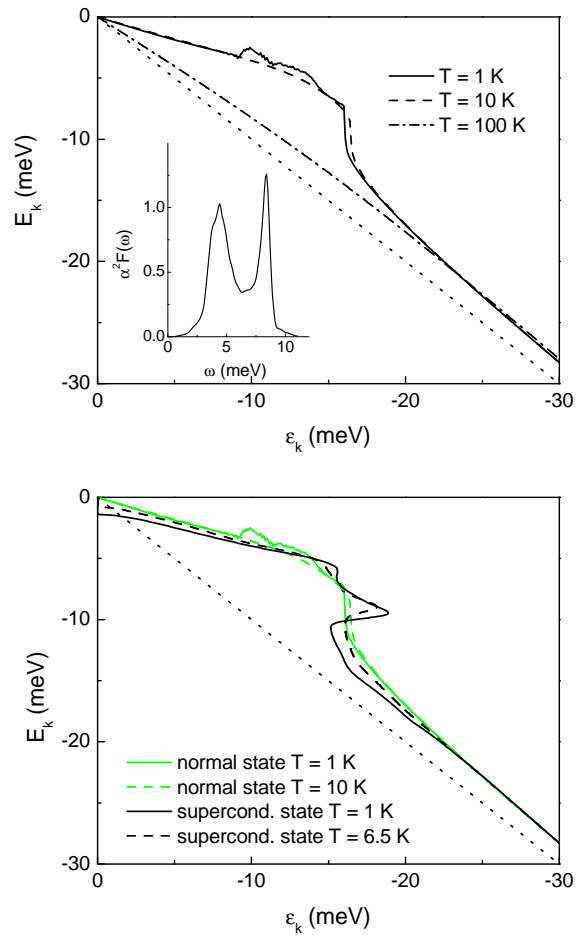}  
  \caption{Top frame: the renormalized energies $E_{\bf k}$ in lead as
    a function
    of the bare band energy $\varepsilon_{\bf k}$; the renormalization
    is due to the electron phonon interaction characterized by the
    electron-phonon spectral density $\alpha^2F(\omega)$ shown in the
    inset. The bare dispersion is shown as the dotted line.
    The other curves
    are for the normal state at $T=1\,$K (solid), $T=10\,$K
    (dashed), and $T=100\,$K (dash-dotted).
    Bottom frame: same as the top frame but for
    the superconducting state at $T=1\,$K (solid)
    and $T=6.5\,$K (dashed) just below $T_c = 7.2\,$K.
    The normal state results
    are shown as gray lines for comparison.
    }
    \label{fig:1}
\end{figure}

So far we have only discussed scattering rates and have emphasized
the differences and similarities between infrared optical absorption
and quasiparticle lifetimes. In ARPES the dressed electronic
dispersion relation is measured and denoted by $E_{\bf k}$. It
is related to the bare band dispersion $\varepsilon_{\bf k}$
through the electron self energy. In the electron-phonon case
with an Einstein phonon spectrum and at $T=0$ (zero temperature)
the self energy reduces to a simple form
\begin{equation}
  \label{eq:8}
  \Sigma(\omega+i0^+) = \frac{\lambda\omega_E}{2}\left[
    \ln\left\vert\frac{\omega_E-\omega}{\omega_E+\omega}\right\vert-
    i\pi\theta(\omega-\omega_E)\right].
\end{equation}
In Eq.~(\ref{eq:8}) $\lambda$ is the electron mass enhancement
parameter defined as $m^\ast/m = 1+\lambda$ where $m^\ast$ is
the renormalized electron mass at the Fermi surface. In terms
of $\alpha^2F(\omega)$, $\lambda = 2\int_0^\infty d\nu\,%
\alpha^2F(\nu)/\nu = 2 A/\omega_E$ for the simplified Einstein
case. The imaginary part of Eq.~(\ref{eq:8}) just gives back
$\tau^{-1}_{qp}(\omega) = -2\Im{\rm m}\Sigma(\omega+i0^+)$ of
Eq.~(\ref{eq:3}). The real part gives the renormalized energies
as solutions of equation
$E_{\bf k} = \varepsilon_{\bf k}+\Re{\rm e}\Sigma(E_{\bf k})$ and
for $\varepsilon_{\bf k}\to 0$ (on the Fermi surface) this gives
$E_{\bf k} = \varepsilon_{\bf k}/(1+\lambda)$: the quasiparticle
mass is simply renormalized to $m^\ast$. Note that in
Eq.~(\ref{eq:8}) the self energy has a logarithmic divergence
at $\omega = \omega_E$ and this leads to a singular structure in
$E_{\bf k}$ at that energy as has been investigated in detail
by Verga {\it et al.}\cite{verga} to which the reader is
referred for details. Here we have chosen to emphasize instead
the nature of the structure in the imaginary part. It is sufficient
to remark that for a finite distribution of phonon energies in
$\alpha^2F(\omega)$ the logarithmic singularity is moderated as
it also is when temperature $(T)$ is included. In the general case,
$\Re{\rm e}\Sigma(\omega+i0^+)$ takes the form
\begin{eqnarray}
  \Re{\rm e}\Sigma(\omega+i0^+) &=&
  -\int\limits_0^\infty\!d\nu\,\alpha^2F(\nu)
  \Re{\rm e}\left[\psi\left(\frac{1}{2}+i\frac{\nu-\omega}{2\pi T}
      \right)\right.\nonumber\\
  &&-\left.\psi\left(\frac{1}{2}-i\frac{\nu-\omega}{2\pi T}
      \right)\right],
  \label{eq:9}
\end{eqnarray}
where $\psi(z)$ is the digamma function. In lead, the electron-phonon
spectral density $\alpha^2F(\omega)$ is well known from tunneling
spectroscopy\cite{tomlinson,carb1,mcmillan,mcmillan1}
as well as from direct calculations. It is shown as an
insert in the top frame of Fig.~\ref{fig:1}.
The phonons extend to $12\,$meV and show a characteristic peak for
transverse and for longitudinal branches.
In the top frame of this figure
we show numerical results in the normal state for $E_{\bf k}$
on the Fermi surface as
a function of the bare band electronic energy $\varepsilon_{\bf k}$
up to $30\,$meV at which point bare (dotted curve) and renormalized
energies are pretty well parallel to each other. The solid
curve applies to the normal state at $T=1\,K$. We see structure in
the renormalized quasiparticle energy $E_{\bf k}$
ranging up to roughly $10\,$meV. The structure
clearly corresponds to the structure in the phonon spectrum (inset).
Beyond this range multi-phonon processes are operative and the
renormalization effects are less.
Further results in the normal state are for $T=10\,$K (dashed)
and $T=100\,$K (dash-dotted).
We see that by the time this last
temperature is reached, which is of the order of the maximum phonon
energy, the thermal effects have smeared out much of the structure.
Note also, that in all three cases
the curves go through zero at $\varepsilon_{\bf k} = 0$ and the
slope of each line out of the origin gives the renormalized effective
mass parameter $(1+\lambda)$ at the temperature $T$.
For the convenience of the reader
we included in the bottom frame of Fig.~\ref{fig:1} results for
the superconducting state at $T=1\,$K (black solid) and at $T=6.5\,$K
(black dashed) just below the critical temperature of lead which
is $T_c = 7.2\,$K. We note that in each case the phonon structure
is somewhat more pronounced than in the corresponding normal
state (shown as gray lines). Also at $\varepsilon_{\bf k}\to
0$ the renormalized energy $E_{\bf k}$ now goes to a finite value
equal to the superconducting gap $\Delta(T)$ at that temperature.
In an ARPES experiment the gap is seen as a shift of the leading
edge of the electron spectral density $A({\bf k},\omega)$,
downward from the chemical potential level.

Before moving on to other possible models for the renormalization
we wish to make an important point about how some of the ARPES
data has been analyzed in the literature.\cite{kaminski,valla,%
  johnson} It is essential to be aware that the bare electron band
energies are not known independently in ARPES experiments.
However, electron momentum change $(k-k_F)$ can be
measured in the particular direction of interest. The bare energy
is then related to this momentum change
through the Fermi velocity $v_F$. The bare value of
$v_F$ denoted $v^0_F$ is then determined from $E_{\bf k}$ under
the assumption that at $E_{\bf k}\simeq 250\,$meV renormalization
effects have become small and are negligible in a first approximation.
This assumption can, however, result in a significant underestimate
of the true mass renormalization $\lambda$ involved as is illustrated
\begin{figure}
  \vspace*{-6mm}
  \includegraphics[width=9cm]{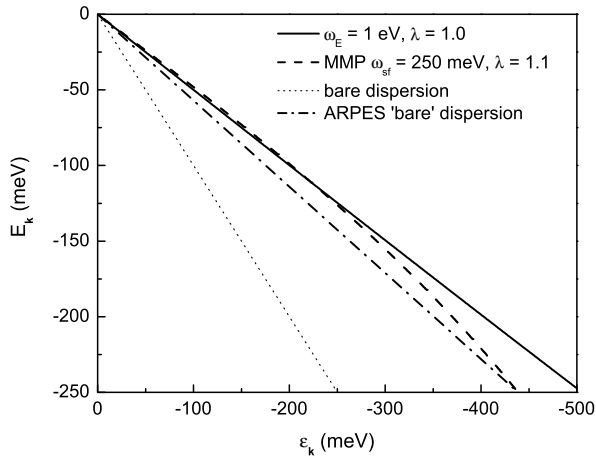}
  \vspace*{-4mm}
  \caption{The renormalized energy $E_{\bf k}$ as a function of
    the bare band energy $\varepsilon_{\bf k}$ for various models
    for the electron-boson exchange spectral density. The
    dotted curve is the bare dispersion relation and is for
    comparison. The solid curve
    is for an Einstein spectrum with $\omega_E = 1\,$eV and
    a mass renormalization factor $\lambda = 1$. The dashed line
    which shows a slight curvature is for a model in which the
    charge carriers are coupled to spin fluctuations described by
    an MMP-form, Eq.~(\protect{\ref{eq:10}}), with a spin fluctuation
    energy $\omega_{sf} = 250\,$meV and a mass renormalization
    $\lambda = 1.1$. The dash-dotted line is a straight line
    construction through $E_{\bf k} = 250\,$meV which is used
    in ARPES experiments to determine the ``bare'' dispersion
    as it is not independently known.
    }
  \label{fig:2}
\end{figure}
in our Fig.~\ref{fig:2}. The solid curve gives $E_{\bf k}$ vs
$\varepsilon_{\bf k}$ for a case where the Einstein frequency
defining $\alpha^2F(\omega)$ is
taken to be very large, $\omega_E = 1\,$eV, compared with the
energies of interest in this figure. The curve is almost a perfect
straight line (there is no structure) with slope giving $(1+\lambda)$
with $\lambda = 1$ by choice. This is to be compared with
the dotted line which gives the bare dispersion
and has slope 1. Applying the construction that the ``ARPES bare
dispersion'' is a straight line going through the renormalized
energy at $E_{\bf k} = 250\,$meV would give the same straight
line as the solid curve and, therefore, we would conclude
that $\lambda = 0$,
i.e.: there is no renormalization. As a second example
we use for the spectral density a form appropriate to a spin
fluctuation model. The form is obtained directly from consideration
of experimental data on the infrared conductivity in the normal
state of the cuprates. It is given by the experimental form
$W(\omega)$ of Eq.~\eqref{eq:10a} which can be adequately
represented by a Lorentzian
\begin{equation}
  \label{eq:10}
  I^2\chi(\omega) \propto W(\omega)
  \propto I^2\frac{\omega}{\omega_{sf}^2+\omega^2}
  \theta(\omega_{\rm max}-\omega).
\end{equation}
(We will refer to this form in the following as MMP-form.)
In Eq.~(\ref{eq:10})
$\omega_{sf}$ is a characteristic spin fluctuation frequency taken for
illustrative purposes $250\,$meV and
$\lambda= 2\int_0^{\omega_{\rm max}}d\nu\,%
I^2\chi(\nu)/\nu = 1.1$ with $\omega_{\rm max} = 400\,$meV. With
this spectral density we get the dashed line in Fig.~\ref{fig:2}
which is to be compared with both, the dotted line (real bare dispersion) and
the dash-dotted line (ARPES constructed `bare' dispersion).
It is clear that the ARPES construction again drastically
underestimates the spin fluctuation mass renormalization
giving a value of $\lambda \sim 0.1$ rather than 1.1. Also,
the curve does not show any identifiable sharp structure, instead it
changes only rather gradually. In order to get the correct
value of $\lambda$ one would need data up to energies higher
than the maximum energy involved in the fluctuation spectrum
that is responsible for the renormalization. For spin fluctuation
theories this energy scale
is set by the magnitude of $J$ of the $t-J$
model and is high of order several hundred meV.\cite{sorella}
The $t-J$ model is widely believed to provide the
appropriate Hamiltonian needed to describe the strongly
correlated charge carriers in the CuO$_2$-planes in the
cuprates. In this instance
ARPES experiments at much higher energies than presently
sampled would be required to get the full value of $\lambda$ involved.

\section{Comparison of theory and experiment in Bi2212}

In the interpretation of their ARPES data some experimentalists
have considered the
possibility that the renormalizations are due to phonons.
We consider here only the specific case of
Bi2212 for which there also exists a recent set of high
accuracy optical conductivity measurements.\cite{tu}
The phonon frequency distribution in this
material has been measured by incoherent inelastic neutron
scattering by Renker {\it et al.}\cite{renker} Such experiments
give directly $F(\omega)$ and a first attempt at a model for an
electron-phonon interaction spectral density $\alpha^2F(\omega)$
can be constructed by multiplying the experimental $F(\omega)$
by a constant adjusted to best fit the ARPES data on the
renormalized quasiparticle energy $E_{\bf k}$ as shown in
\begin{figure}
  \includegraphics[width=9cm]{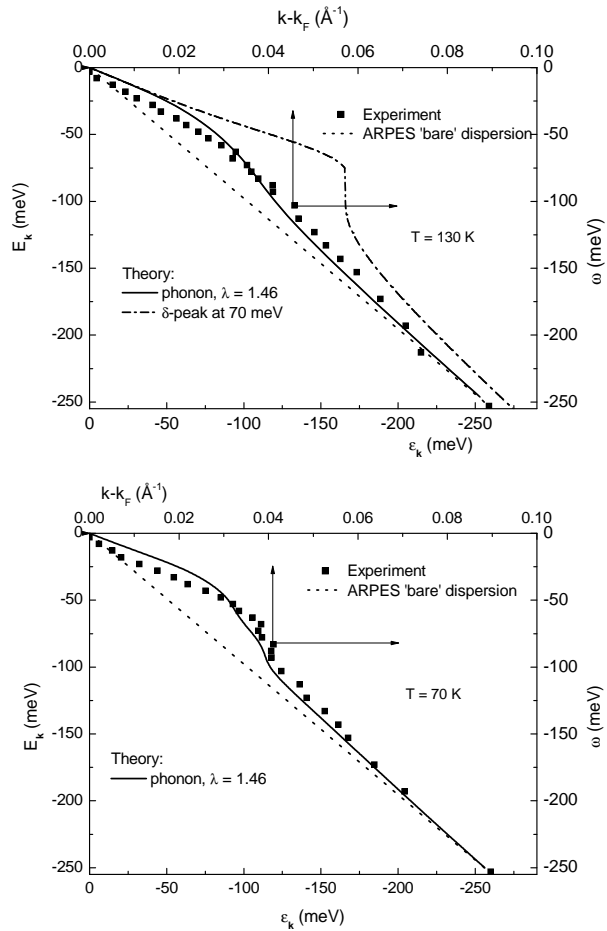}
  \vspace*{-5mm}
  \caption{Top frame: the renormalized energy $E_{\bf k}$ as a function of
    bare band energy $\varepsilon_{\bf k}$ using an electron-phonon
    spectral density $\alpha^2F(\omega)$ to describe the coupling.
    The solid squares are the data of Johnson {\it et. al.}
    \protect{\cite{johnson}} and the solid curve is our fit. The
    full phonon spectral density revealed by inelastic neutron
    scattering\protect{\cite{renker}} was used with the area
    under $\alpha^2F(\omega)$ adjusted to get a mass renormalization
    parameter $\lambda = 1.46$. In the theory at $k-k_F = 0.089\,$
    \AA$^{-1}$ $\varepsilon_{\bf k} = 256\,$meV
    while the value obtained from the
    ARPES construction is nearly the same, $253\,$meV. The
    curve is for $T=130\,$K. The dash-dotted curve is for a
    $\delta$-function model for $\alpha^2F(\omega)$ at $70\,$meV
    and $\lambda = 0.98$. The bottom frame
    applies to the lower temperature $T=70\,$K and is in the
    superconducting state.
    }
  \label{fig:3}
\end{figure}
Fig.~\ref{fig:3}. The data\cite{johnson} at $T=130\,$K are shown
in the top frame and at $T=70\,$K in the bottom frame as solid squares.
On the top horizontal scale we show momentum differences $k-k_F$
in \AA$^{-1}$ while on the bottom horizontal scale
we have the value corresponding to our bare band energy
$\varepsilon_{\bf k}$ which implies a theoretical bare energy of
$256\,$meV at $k-k_F = 0.089\,$\AA$^{-1}$.
The fit in both cases was obtained with a value
of $\lambda = 1.46$. We see that the resulting theoretical results
(solid curves) fit reasonably well the experimental data. Here, no
attempt has been made to get a best fit by varying the shape of
the underlying spectral density $\alpha^2F(\omega)$.
The reader is referred
to the work by Verga {\it et al.}\cite{verga} for an alternative
approach to the analysis of ARPES data
  for the material La$_{2-x}$Sr$_x$CuO$_4$ (LSCO). These authors do attempt
to get the shape of the spectrum from the data itself and we will
return to this issue later. Two final comments on Fig.~\ref{fig:3}.
The dotted curve which is the ARPES construction for the bare
dispersion gives at $k-k_F = 0.089\,$\AA$^{-1}$ an
$\varepsilon_{\bf k} = 253\,$meV which is nearly the same as the
theoretical value of $256\,$meV.
This is to be expected since phonon renormalization
effects are small at $250\,$meV and beyond.
Secondly, we wish to point out
that the data at $T=70\,$K is in the superconducting state. The
formalism needed to treat superconductivity
is more complicated than that for the normal state
which we have sketched,
but the ideas and concepts are the same and no details are
included here. The reader is referred to an extensive
existing literature.\cite{carb1,mars1} Here it will be sufficient to point
out that the value of the critical temperature for superconductivity
in a $d$-wave superconductor is
determined most directly by a different function
that which comes into the mass
renormalization channel. For an $s$-wave isotropic superconductor
this issue does not arise. The quasiparticle $\alpha^2F(\omega)$
would also determine superconductivity and one could ask
if ARPES data on $\lambda$ are indeed consistent with the observed
superconductivity. But in the cuprates the superconducting gap
has $d$-wave symmetry. In this case it is the projection of
the quantity  $\left\vert g_{{\bf k},{\bf k}',\nu}\right\vert^2%
\delta(\omega-\omega_\nu({\bf k}-{\bf k}'))$ onto the $d$-wave
channel which determines the spectral function which is most directly
involved in superconductivity (rather than its projection on the
$s$-channel).
ARPES does not measure this
projection separately and so this technique, stricktly speaking,
remains silent on the issue of mechanism
for superconductivity even if phonons should be the main
cause of renormalization of the electron dispersion curves.

Returning to the top frame of Fig.~\ref{fig:3} the dash-dotted
curve which does not agree well with the data is shown for
comparison. It is based on a model electron-phonon spectral
density $\alpha^2F(\omega)$ in which the coupling is entirely
to a single frequency fixed at $70\,$meV\cite{lanzara,bogdanov}
with a mass enhancement
factor $\lambda = 0.98$ chosen to get a critical temperature
of $T_c = 91\,$K when $s$- and $d$-channel spectral densities are
taken to be the same. Such a model does not agree well with experiment
and when it is used, in the superconducting state at $T=70\,$K,
the disagreement is even worse. The $\delta$-function model
corresponds to an extreme in which it is assumed that the
electron-phonon coupling is to a single mode.
This is unlikely to be the case. A model based on the entire
phonon spectrum is more realistic although, as we said
before, $\alpha^2F(\omega)$ does not need to have the same
shape as $F(\omega)$. Special modes could be weighted more heavily
in $\alpha^2F(\omega)$ as compared to $F(\omega)$.

\section{Optical conductivity}
\label{sec:4} 

We can impose a second constraint on renormalization effects by
considering the infrared conductivity. There exists much data on
optical conductivity in the cuprates and these have given us
considerable insight into the inelastic scattering involved. The
conductivity $\sigma(\omega)$ at $T=0$ (to remain simple) in the
normal state also follows in a first approximation
from the knowledge of the electron self
energy:\cite{mars1}
\begin{equation}
  \label{eq:11}
  \sigma(\omega) = \frac{\Omega^2_p}{4\pi}\int\limits_0^\omega d\nu\,
  \frac{1}{i\tau^{-1}_{\rm imp}-\Sigma(\nu)-\Sigma(\omega-\nu)}.
\end{equation}
More complicated formulas exist at any finite temperature $T$
and/or for the superconducting case and including vertex corrections
which we will not
repeat here for the sake of brevity. They can be found in many
places including our own work. Since here we will fit the
electron-boson exchange density to optical data directly, the
resulting form can be thought of as including higher
order corrections.

In Eq.~(\ref{eq:11})
$\tau^{-1}_{\rm imp}$ is the elastic impurity scattering rate
while the imaginary part of $\Sigma(\omega)$ deals with the corresponding
inelastic scattering.
What is most often measured in optical experiments, is the
reflectance $R(\omega)$ as a function of $\omega$. This is
\begin{figure}
  \vspace{-5mm}
  \includegraphics[width=9cm]{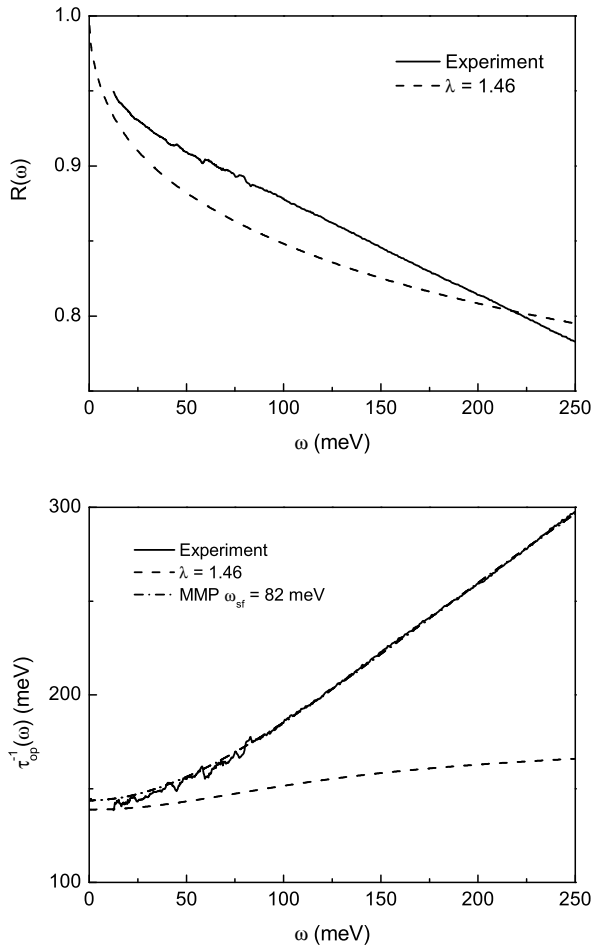}
  \vspace*{-4mm}
  \caption{Top frame: the reflectance for Bi2212 in the infrared up to
    $\omega = 250\,$meV. The solid curve presents the data of
    Tu {\it et al.}\protect{\cite{tu}} The dashed curve is for the
    phonon spectrum and a $\lambda = 1.46$. The temperature is
    $T=295\,$K. The bottom frame shows the derived optical scattering rate
    $\tau^{-1}_{op}(\omega)$ as a function of $\omega$. The solid
    curve presents the data of Tu {\it et al.}\protect{\cite{tu}}
    The dashed curve is a phonon based result
    which does not give enough
    variation with increasing $\omega$. By contrast, the
    dash-dotted curve fits well and is based on a simple spin
    fluctuation spectrum of MMP-form with $\omega_{sf} = 82\,$meV
    and $\lambda = 1.79$.
    }
  \label{fig:4}
\end{figure}
shown in the top frame of Fig.~\ref{fig:4} for Bi2212. The solid curve
is the data of Tu {\it et al.}\cite{tu} for room temperature
$T=296\,$K.
The dashed curve was obtained with the same neutron based
  electron-phonon interaction spectrum having a $\lambda = 1.46$
  as we found from our ARPES analysis.
  It provides a poor fit as is illustrated, even more sharply,
  in the bottom frame of Fig.~\ref{fig:4} which deals with
  optical scattering rates.
It has now become standard procedure for experimentalists to
extract from their reflectance data the real and imaginary part of
the optical conductivity and to construct from this information the
optical scattering rate $\tau^{-1}_{op}(\omega)$. The solid line
in the bottom frame is the data. The two other
curves are theoretical and the dashed is based on the neutron 
$\alpha^2F(\omega)$ with 
$\lambda = 1.46$. It is clear that it does not
agree with the data, particularly as $\omega$ increases. This
defect remains whatever
value of $\lambda$ is used. The fundamental problem is that the
phonon spectrum cuts off too soon as a function of frequency to fit the
optics and, consequently, the calculated $\omega$-dependence of
$\tau^{-1}_{op}(\omega)$ is much too flat.
Any boson mechanism that fits the data will need to have an
energy scale that extends to order of a few $100\,$meV or so. This is
consistent with a spin NAFFL\cite{pines1,pines2} or a MFL
model\cite{varma1,varma2,varma3} but is not compatible with
phonons. It is important to note that this limitation of a
phonon mechanism is independent of any difference there might be
in the size of $\alpha^2_{tr}F(\omega)$ as compared with
$\alpha^2F(\omega)$. Both functions cut off at the maximum phonon
energy ($\omega_{\rm max} = 80\,$meV in case of Bi2212) and this
cannot give agreement with the optics. On the other hand, a spin
fluctuation model provides a natural explanation because it
involves a higher boson energy scale of the order $J$
of the $t-J$-model as previously noted.\cite{sorella}
This is demonstrated in the lower frame of Fig.~\ref{fig:4} by the
dash-dotted curve which
is for an MMP-form with $\omega_{sf} = 82\,$meV and
$\lambda = 1.79$. It fits
the experiment almost perfectly with no need for any adjustment
of any kind. (Deviations from $\omega_{sf} = 82\,$meV within $\pm 5\,$meV will
change the quality of this fit only marginally.)
 It is interesting to note in closing this
  discussion that in the case of electron-phonon interaction
  the same value for $\lambda$ is required to get a reasonable
  fit of the ARPES data and to get agreement at least in the
  low energy regime with the optical data. This is in conflict
  with the findings by Kuli\'c\cite{no15} that in the phonon
  assisted case $\alpha^2_{tr}F(\omega)$ should be significantly
  smaller than $\alpha^2F(\omega)$. 

\begin{figure}
  \vspace*{-6mm}
  \includegraphics[width=9cm]{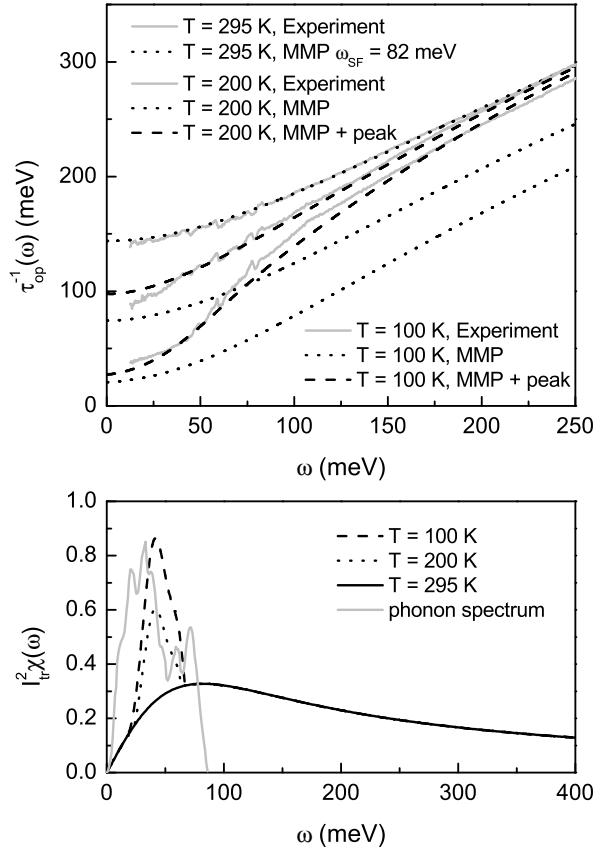}
  \caption{Top frame: optical scattering rates for Bi2212 at $T=295\,$K,
    $T=200\,$K, and $T=100\,$K. All are in the normal state. The
    gray solid lines
    are experiment of Tu {\it et al.}\protect{\cite{tu}}, dotted
    are theory based on an MMP-form for the charge
    carrier-exchange boson spectral density, and the dashed
    have in addition coupling to an optical resonance at $43\,$meV.
    The bottom frame shows the derived spectral density based on 
    our well defined
    procedure to invert optical data. The $T=295\,$K spectrum has
    no resonance and is a pure MMP-form with $\omega_{sf} = 82\,$meV
    (solid line). The
    other two spectra are modified through the addition of coupling
    to a resonance mode at $43\,$meV. Also shown is the phonon
    spectrum (gray solid curve).\protect{\cite{renker}}
    }
  \label{fig:5}
\end{figure}

In the top frame of Fig.~\ref{fig:5} we compare theory with experiment
at three
temperatures, namely $T=100\,$K, $T=200\,$K, and $T=295\,$K. The data
on the optical scattering rate was obtained by Tu
{\it et al.}\cite{tu} from their infrared measurements
in Bi2212 and is denoted by gray solid lines
(top to bottom is decreasing temperature). The black dotted lines are
theory obtained from an MMP-form given by Eq.~%
(\ref{eq:10}) with $\omega_{sf} = 82\,$meV. The $T=295\,$K
curve repeats the fit shown in the bottom frame of Fig.~\ref{fig:4}.
The same spectral density, however, does not fit
well the $T=200\,$K and $T=100\,$K data (dotted lines). To get
agreement with these data sets it is necessary to augment the MMP-%
form by adding coupling to an additional optical resonance
at $43\,$meV (this gives the dashed curves).
This is the frequency at which
inelastic neutron scattering experiments have revealed a strong
peak in the magnetic susceptibility at momentum $(\pi,\pi)$
in the two dimensional Brillouin zone of the CuO$_2$ plane,%
\cite{fong} in the superconducting state. The first report of such
a resonance in neutron scattering was by J.~Rossat-Mignot
{\it et al.} for YBa$_2$Cu$_3$O$_{6+\delta}$.\cite{ros}
We have already described our method
for obtaining $I^2_{tr}\chi(\omega)$,\cite{mars4,mars5,schach4,schach7,%
schach5,schach8,schach9} thus it is not necessary to give
details here except to mention that Tu {\it et al.}\cite{tu}
also get this peak using a slightly different method.
We present final results for our model $I_{tr}^2\chi(\omega)$
in the bottom frame of Fig.~\ref{fig:5}.
While no peak at $43\,$meV is observed at $T=295\,$K (solid line)
one shows up at lower temperatures, dotted line for $T=200\,$K and
a somewhat bigger peak at $T=100\,$K (dashed curve). At present, neutron
scattering does not reveal a spin resonance at these temperatures.
 Returning to the top frame
of Fig.~\ref{fig:5} we stress that the introduction of a
resonance peak in addition to the MMP-form is
responsible for the observed rapid rise around $50\,$meV in
the optical scattering rate at $T=100\,$K and also, although
less obvious, at $T=200\,$K. The model spectral density fits the
data remarkably well.
Such optical resonances have been seen before
in YBa$_2$Cu$_3$O$_{6.96}$\cite{schach4} as well as in
Tl compounds\cite{schach5,schach9} in the superconducting state of
  optimally doped samples, and have their equivalent
in spin resonances observed by inelastic neutron scattering.%
\cite{fong,ros,dai,he}
Here the high temperature optical resonances in Bi2212 are seen
for the first time in the
normal state above $T_c$, as reported by Tu {\it et al.}\cite{tu}
Their exact microscopic origin, however,
cannot be deduced from optical data alone. It could be that the
$43\,$meV neutron spin resonance forms above $T_c$ in this
optimally doped material. Alternatively one might argue that this may be
a phonon contribution (see gray solid curve in the bottom frame
of Fig.~\ref{fig:5}) to the total electron-boson
transport spectral density. If this were the case, however, we would
not expect the peak of Fig.~\ref{fig:5} (bottom frame)
to show significant temperature dependence
in this temperature regime. Also, the phonon peak in the bottom
frame of Fig.~\ref{fig:5} has more weight at low $\omega$ than
is indicated in optics. In terms of the area under the
spectral density $I^2_{tr}\chi(\omega)$ we note that the peak itself
contributes less than about 15\% of the total weight (at
$T=100\,$K), the rest comes from the MMP-form. While the
issue of the origin of the peak is important, here we concentrate
instead on the consequences of its existence for the ARPES data.
The possibility of coupling to a resonance mode has been discussed
in the past in connection with such data.\cite{norm1,norm2,norm3}

\begin{figure}
  \includegraphics[width=9cm]{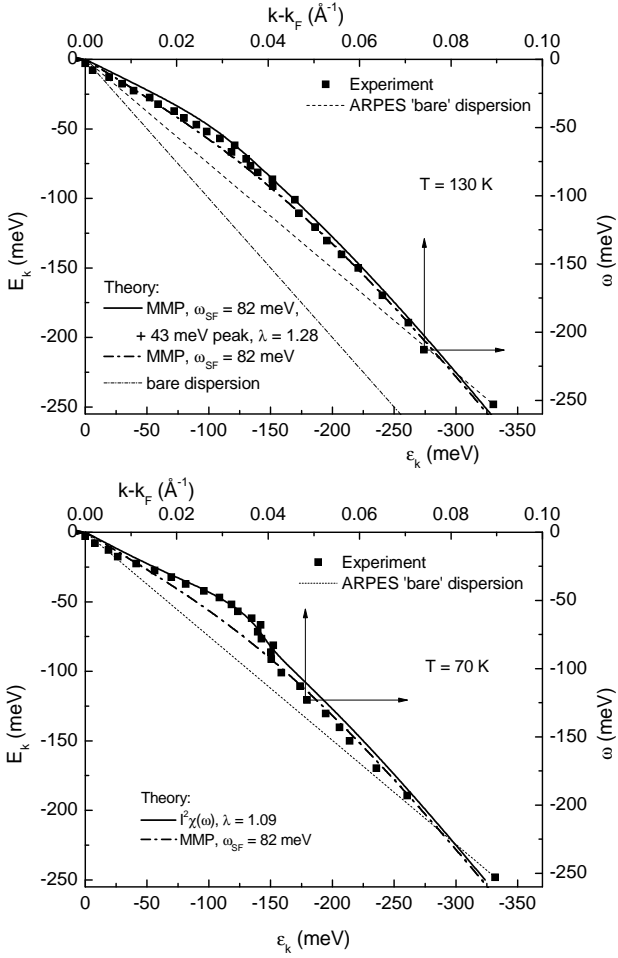}
  \caption{Top frame: the renormalized quasiparticle dispersion $E_{\bf k}$
    as a function of the bare band energy $\varepsilon_{\bf k}$ in
    meV (light dash-dotted).
    The solid curve is based on the spin fluctuation
    spectral density with resonance peak
    obtained from conductivity data (Fig.~\protect{%
      \ref{fig:5}}) scaled down to a mass renormalization of
    $\lambda = 1.28$, otherwise left unmodified. The
    heavy dash-dotted curve is for comparison and has no resonance peak.
    The light dashed
    straight line is for comparison and is the ARPES ``bare
    dispersion'' obtained by making the curve go through experiment
    at $253\,$meV. This corresponds to a theoretical
    $\varepsilon_{\bf k} = 327\,$meV at $k-k_F=0.089\,$\AA$^{-1}$,
    much greater than the ARPES `bare' value.
    This frame applies to $T=130\,$K and is in the
    normal state. The bottom frame is for $T=70\,$K in the superconducting
    state.
    }
  \label{fig:6}
\end{figure}

In Fig.~\ref{fig:6} we compare our theoretical results for the
renormalized quasiparticle energies with the ARPES experimental
data (solid squares).
The theory (solid curve) is based on the optics derived spectral density
(see Fig.~\ref{fig:5}, bottom frame). The top frame is for $T=130\,$K
in the normal state and the bottom frame is for $T=70\,$K in the
superconducting state. For the top frame we renormalized downward
the $I^2_{tr}\chi(\omega)$ to account for the expectation that in
a spin fluctuation mechanism the quasiparticle electron-boson
interaction spectral density should be smaller than its transport
counterpart. To get the fit seen in the figure we used $\lambda =
1.28$. A smaller value of $\lambda = 1.09$ was used in
the bottom frame. We see that we can get an equally good fit to
ARPES with spin fluctuations including the $43\,$meV resonance as we
did with phonons. This new fit (see Fig.~\ref{fig:6}),
however, has the advantage that it can
equally well explain the optical data while this is impossible with
phonons. It is also clear from the calculations that the transport
electron-boson interaction spectral density is larger than the
quasiparticle spectral density by a factor of about two
as is expected in theories of the NAFFL model. It should be
acknowledged, however, that part of the difference in the
value of $\lambda$ needed to fit ARPES data as compared with
optics reflects anisotropies in the quasiparticle renormalizations
over the Brillouin zone.

Also included in the two frames of Fig.~\ref{fig:6} are our
results for a $E_{\bf k}$ (heavy dash-dotted)
when the resonant peak in the spectral
density is left out of the calculations. We see that the pronounced
structure at $k-k_F\sim 0.04\,$\AA$^{-1}$ seen in the superconducting
state is not reproduced in this case. If we used
instead the MFL model, i.e.: $I^2\chi(\omega) = I^2\theta(\omega_{max}%
-\omega)$, the deviation from the data would be worse and
even clearly seen
in the normal state. The resonance peak, absent in the MMP-form and
in the MFL model, is what
gives the observed structure. Also shown in the figure is the
the bare dispersion relation (light dash-dotted curve) on which
our calculations are based. This is quite different from the
ARPES derived `bare' dispersion, the light dashed curve, obtained by
making the curve go through the experimental point at $253\,$meV
on the assumption that the renormalization by boson-exchange
interaction has ended at this energy. Reference to the difference
between the theoretical bare dispersion (light dash-dotted)
and the dressed curve
(solid line) shows clearly that this is not the case in a spin
fluctuation model.

It is interesting to note the similarities and differences
between our work and that of Verga {\it et al.}\cite{verga}
An important difference is that these authors invert the
ARPES data directly from which they obtain a model
electron-boson spectral density. Instead, we have
inverted the optical data which serves as a second constraint
on the microscopics not considered by Verga {\it et al.}
While they invert data in LSCO at three
different doping levels $x$, they, nevertheless, find a spectrum
which has a peak which is, however, not as prominent as
either the spin resonance peak or the phonon peak in the
bottom frame of our Fig.~\ref{fig:5}. Their peak could
simply be a peak of the MMP-form as in the solid black
curve in the bottom frame of Fig.~\ref{fig:5}. Nothing
definite can be concluded. However, we point out that they
do find tails at higher energies
($\omega > 100\,$meV), as we have, which rather support
spin fluctuations and not phonons.
In this paper we consider in addition to ARPES
optical data and from this we conclude more forcefully that
high energy tails extending up to several hundred meV exist
in the boson assisted spectral functions describing the
electron boson interaction in Bi2212.

Whatever the mechanism by which the quasiparticles are renormalized
the $(11)$-component of the $2\times 2$ Nambu Green's function
in the superconducting state can be expressed in the general form%
\cite{mars6}
\begin{eqnarray}
  \label{eq:12}
  G_{11}({\bf k},\omega) =
  \frac{\omega Z_{\bf k}(\omega)+\varepsilon_{\bf k}}
  {\omega^2 Z^2_{\bf k}(\omega)-\varepsilon^2_{\bf k}
    -\tilde{\Delta}^2_{\bf k}(\omega)},
\end{eqnarray}
where the renormalized frequency $\tilde{\omega}(\omega)\equiv
\omega Z(\omega)$. The inverse quasiparticle lifetime is defined
as twice the imaginary part of the poles of $G_{11}({\bf k},\omega)$
for a given momentum label {\bf k}. For $\omega = E_{\bf k}-
i\Gamma_{\bf k}$ we obtain\cite{mars6}
\begin{equation}
  \label{eq:13}
  \Gamma_{\bf k} = \frac{E_{\bf k}Z_{{\bf k}2}(E_{\bf k})}
    {Z_{{\bf k}1}(E_{\bf k})}-\frac{\tilde{\Delta}_{{\bf k}1}
        (E_{\bf k})\tilde{\Delta}_{{\bf k}2}(E_{\bf k})}
      {E_{\bf k}Z^2_{{\bf k}1}(E_{\bf k})},
\end{equation}
where the indices 1 and 2 refer to the real and imaginary
  part respectively,
and the corresponding renormalized energy is a solution of
\begin{equation}
  \label{eq:14}
  E_{\bf k} = \sqrt{\frac{\varepsilon^2_{\bf k}+
      \tilde{\Delta}^2_{{\bf k}1}(E_{\bf k})}
    {Z^2_{{\bf k}1}(E_{\bf k})}}.
\end{equation}
For the normal state $(\tilde{\Delta}_{\bf k}(\omega) = 0)$ these
expressions reduce to those previously given in the limit of weak
scattering with the modification that $\Gamma$ is renormalized by
a mass enhancement factor $(1+\lambda)$.

\begin{figure}
  \vspace*{-5mm}
  \includegraphics[width=9cm]{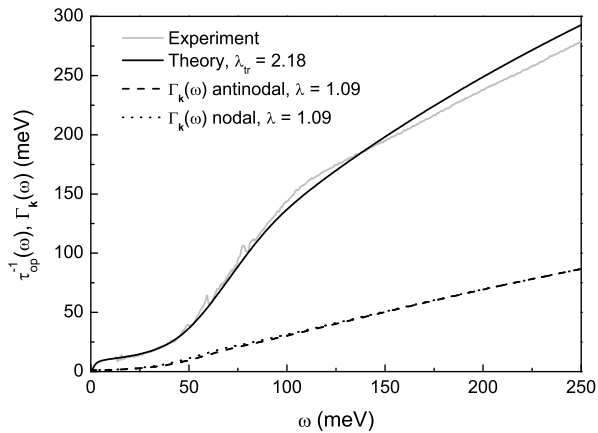}
  \caption{Comparison between scattering rates as a function
    of frequency $\omega$ obtained from optics and from ARPES.
    The gray solid line is the experimental
    optical scattering rate at $T=80\,$K obtained from the work of
    Tu {\it et al.}\protect{\cite{tu}} while the black solid curve
    is our theoretical result based on the
    electron-spin fluctuation spectral density $I^2_{tr}\chi(\omega)$
    shown in the bottom frame of Fig.~\protect{\ref{fig:5}}
    (but using data for $T=80\,$K). The other two curves are
    theoretical results for the
    imaginary part of the quasiparticle self energy.
    }
  \label{fig:7}
\end{figure}

In Fig.~\ref{fig:7} we emphasize once again the difference obtained
between quasiparticle scattering rates and its optical counterpart.
It applies to $T=80\,$K for Bi2212 in the superconducting state.
The gray solid curve represents the data of Tu {\it et al.}%
\cite{tu} for the optical scattering rate and the solid line our
theoretical fit to this data which also determines the transport
spectral density $I^2_{tr}\chi(\omega)$. The corresponding value
of $\lambda_{tr}$ is 2.18. Consideration
of the ARPES data on quasiparticle renormalizations shows us that the
equilibrium spectral density $I^2\chi(\omega)$ must be considerably
smaller. To calculate the imaginary part of
$\Gamma(\omega)$ of Eq.~(\ref{eq:13}) we have reduced
$I^2_{tr}\chi(\omega)$ by a factor of two but without changing
its shape. This is done to get reasonable agreement with ARPES shown in
  Fig.~\ref{fig:6}. The results
are shown as dashed and dotted lines for the antinodal and nodal
directions respectively. Not only is the shape of the resulting
scattering rate different  from the frequency variation obtained
in the optical case but its magnitude is also different reflecting
the difference in magnitude of $I^2_{tr}\chi(\omega)$ and
$I^2\chi(\omega)$.

\section{Conclusion}

We have reconsidered the interpretation of the quasiparticle
renormalizations observed in high resolution
ARPES data for the specific case of Bi2212. We confirm that the
observed energy scale is compatible with coupling to phonons.
We find, however, that present data and, in particular, the
way they are analyzed might well significantly underestimate
renormalization due to other boson exchange mechanisms distinct
from phonon exchange and for which the energy scale is larger. As an
example this would be the case in a spin fluctuation exchange
mechanism as is envisaged in the NAFFL model where the energy
scale for the spin fluctuations is set by $J$ of the $t-J$ model
and is consequently very high.

If in addition to ARPES, optical data is also considered, it is
found that an explanation in terms of phonons becomes less
tenable. Of course, there are many complications and these
make a definitive interpretation impossible. What is clear,
however, is that transport properties and in particular the
observed infrared conductivity cannot be understood as due to
the electron-phonon interaction. This holds even if modifications
are introduced to account for the reduced screening present
in the cuprates because of the low electron density and also
the increased effects of correlations.  The main point is that
the underlying charge carrier-phonon spectral density is cut off
at the maximum phonon frequency and this feature is in conflict
with optical data. By contrast, spin fluctuations give a
natural explanation of the data. Our analysis of the
data of Tu {\it et al.}\cite{tu} also provides a natural
explanation of the ARPES data. What we find is that an optical
resonance is present in the infrared data even in the normal state
above $T_c$ and this resonance can explain the energy scale
seen in ARPES. (Below $T_c$ this optical resonance
has a spin resonance equivalent observed by Fong {\it et al.}\cite{fong}
using inelastic neutron scattering.)
The microscopic nature of the optical resonance
is not entirely known. It could have its origin in spin
fluctuations theories and be the spin one resonance observed
in inelastic polarized neutron scattering experiments, or
could even contain a phonon contribution over and beyond the spin
fluctuation contribution.

Our analysis of both ARPES and optical data together has served to
emphasize once more the differences between the equilibrium
electron-boson exchange spectral density $\alpha^2F(\omega)$
and its transport counterpart $\alpha^2_{tr}F(\omega)$ be it
for phonon exchange or spin fluctuations. The data indicates that
these two quantities differ in magnitude by roughly a factor
of two or so. This is understood to arise from the fact that
transport scattering rates emphasize more strongly backward
collisions than do equilibrium rates. While detailed and elaborate
calculations beyond the scope of this paper would be required
to pin down the expected  differences quantitatively, present
ideas about reduced screening in the cuprates lead directly to
the expectation that phonons will be much less important in
transport than for equilibrium properties while ideas about the
NAFFL model lead to opposite expectations. Spin fluctuations
dominate the transport while at the same time could be less
important in determining quasiparticle properties. This
agrees well with the available combined set of optical and
ARPES data in Bi2212.

\begin{acknowledgments}
Research supported by the Natural Sciences and Engineering
Research Council of Canada (NSERC) and by the Canadian
Institute for Advanced Research (CIAR). We thank Dr.
T.\ Valla for discussions and for making his data available to us.
\end{acknowledgments}


\begin{thebibliography}{99}
  \bibitem{kaminski}A.~Kaminski, J.~Mesot, H.~Fretwell, J.C.~%
    Campuzano, M.R.~Norman, M.~Randeria, H.~Ding, T.~Sato,
    T.~Takahashi, T.~Mochiku, K.~Kadowaki, and H.~Hoechst,
    \prl {\bf 84}, 1788 (2000).
  \bibitem{valla}T.~Valla, A.V.~Fedorov, P.D.~Johnson, J.~Xue,
    K.E.~Smith, and F.J.~DiSalvo, \prl {\bf 85}, 4759 (2000).
  \bibitem{johnson}P.D.~Johnson, T.~Valla, A.V.~Fedorov, Z.~Yusof,
    B.O.~Wells, Q.~Li, A.R.~Moodenbaugh, G.D.~Gu, N.~Koshizuka,
    C.~Kendziora, Sha Jian, and D.G.~Hinks, \prl {\bf 87}, 177007 (2001).
  \bibitem{kaminski1}A.~Kaminski, M.~Randeria, J.C.~Campuzano,
    M.R.~Norman, H.~Fretwell, J.~Mesot, T.~Sato, T.~Takahashi,
    and K.~Kadowaki, \prl {\bf 86}, 1070 (2001).
  \bibitem{lanzara}A.~Lanzara, P.V.~Bogdanov, X.J.~Zhou, S.A.~Kellar,
      D.L.~Feng, E.D.~Lu, T.~Yoshida, H.~Eisaki, A.~Fujimori,
      K.~Kishio, J.-I.~Shimoyama, T.~Noda, S.~Uchida,
      Z.~Hussain, and Z.-X. Shen, Nature (London) {\bf 412},
      510 (2001).
  \bibitem{bogdanov}P.V.~Bogdanov, A.~Lanzara, S.A.~Kellar, X.J.~%
    Zhou, E.D. Lu, W.J.~Zheng, G.~Gu, J.-I.~Shimoyama, K.~Kishio,
    H.~Ikeda, R.~Yoshizaki,
    Z.~Hussain, and Z.-X. Shen, \prl {\bf 85}, 2581 (2000).
  \bibitem{valla1}T.~Valla, A.V.~Fedorov, P.D.~Johnson,
    B.O.~Wells, S.L.~Hulbert, Q.~Li, G.D.~Gu, N.~Koshizuka,
    Science {\bf 285}, 2110 (1999).
  \bibitem{valla2}T.~Valla, A.V.~Fedorov, P.D.~Johnson, Q.~Li,
    G.D.~Gu, and N.~Koshizuka, \prl {\bf 85}, 828 (2000).
  \bibitem{mars6}F.\ Marsiglio and J.P.\ Carbotte, in Handbook on
    Superconductivity: Conventional and Unconventional, eds.
    K.H.\ Bennemann and J.B.\ Ketterson (Springer, Berlin, in print).
  \bibitem{hardy}W.N.~Hardy, D.A.~Bonn, D.C.~Morgan, R.~Liang,
    and K.~Zhang, \prl {\bf 70}, 3999 (1993).
  \bibitem{shen}Z.-X.~Shen, D.S.~Dessau, B.O.~Wells, D.M.~King,
    W.E.~Spicer, A.J.~Arko, D.~Marshall, L.W.~Lombardo, A.~Kapitulnik,
    P.~Dickinson, S.~Doniach, J.~DiCarlo, A.G.~Loeser, and C.H.~Park,
    \prl {\bf 70}, 1553 (1993).
  \bibitem{wollm}D.H.~Wollman, D.A.~Van Harlingen, W.C.~Lee,
    D.M.~Ginsberg, and A.J.~Leggett, \prl {\bf 71}, 2134 (1993).
  \bibitem{tsuei}C.C.~Tsuei, J.R.~Kirtley, C.C.~Chi, LockSeeYu-Jahnes,
    A.~Gubta, T.~Shaw, J.Z.~Sun, and M.B.~Ketchen, \prl {\bf 73},
    593 (1994).
  \bibitem{tomlinson}P.~Tomlinson and J.P.~Carbotte, \prb {\bf 13},
    4738 (1976).
  \bibitem{carb2}J.P.~Carbotte, in {\it Anisotropy Effects in
      Superconductors}, ed. H.W.~Weber, Plenum (New York, 1977),
    p. 183.
  \bibitem{no15}M.L.\ Kuli\'c, Phys.\ Rep. {\bf 338}, 1 (2000).
  \bibitem{zeyher}R.~Zeyher and M.L.~Kuli\'c, \prb {\bf 53},
    2850 (1996); {\bf 54}, 8985 (1996).
  \bibitem{kulic}M.L.~Kuli\'c and R.~Zeyher, \prb {\bf 49},
    4395 (1994); Physica C {\bf 199-200}, 358 (1994);
    {\bf 235-240}, 358 (1994).
  \bibitem{weger}M.~Weger, B.~Barbelini, and M.~Peter,
    Z.\ Phys.\ B {\bf 94}, 387 (1994).
  \bibitem{weger1}M.~Weger, M.~Peter, and L.P.~Pitaevskii,
    Z.\ Phys. B {\bf 101}, 573 (1996).
  \bibitem{kulic1}O.V.\ Danylenko, O.V.\ Dolgov, M.L.\ Kuli\'c, and
     V.\ Oudovenko, Euro.\ Phys.\ Jour.\ B-Cond.\ Matter {\bf 9},
     201 (1999).
  \bibitem{kulic2}M.L.~Kuli\'c and O.V.~Dolgov, in {\it
   High Temperature Superconductivity}, ed.: S.~Barnes,
   J.~Ashkenazi, J.~Cohn, and F.~Zuo, AIP Conference
   Proceedings {\bf 483}, 63 (1999).
  \bibitem{sorella}S.~Sorella, G.B.~Martins, F.~Becca, C.~Gazza,
   L.~Capriotti, A.~Parola, and E.~Dagotto, \prl {\bf 88},
   117002 (2002).
  \bibitem{pines1}A.J.~Millis, H.\ Monien, and D.\ Pines,
     \prb {\bf 42}, 167 (1990).
  \bibitem{pines2}P.\ Monthoux and D.\ Pines, \prb {\bf 47},
     6069 (1993); \prb {\bf 49}, 4261 (1994); \prb {\bf 50},
     16015 (1994).
  \bibitem{varma1}C.M.\ Varma, Int.\ J.\ Mod.\ Phys. {\bf 3},
    2083 (1989).
  \bibitem{varma2}P.B.\ Littlewood, C.M.\ Varma, S.\ Schmitt-Rink,
    and E.~Abrahams, \prb {\bf 39}, 12371 (1989).
  \bibitem{varma3}C.M.~Varma, P.B.~Littlewood, S.~Schmitt-Rink,
    E.~Abrahams, and A.E.~Ruckenstein, \prl {\bf 63}, 1996 (1989);
    {\it ibid.} {\bf 64}, 497 (1990).
  \bibitem{puchkov}A.V.\ Puchkov, D.N.\ Basov, and T.\ Timusk,
    J.\ Phys.: Condens.\ Matter {\bf 8}, 10049 (1996).
  \bibitem{timusk}T.~Timusk and B.~Statt, Rep.\ Prog.\ Phys.
    {\bf 62}, 61 (1999).
  \bibitem{tu}J.J.\ Tu, C.C.\ Homes, G.D.\ Gu, D.N.\ Basov,
    and M.\ Strongin, \prb {\bf 66}, 144514 (2002).
  \bibitem{tomlinson1}P.~Tomlinson and J.P.~Carbotte, Can.\ %
    J.\ Phys. {\bf 55}, 751 (1977).
  \bibitem{leung}H.K.~Leung, F.W.~Kus, N.~McKay, and
    J.P.~Carbotte, \prb {\bf 16}, 4358 (1977).
  \bibitem{allen}P.B.~Allen, \prb {\bf 3} 305 (1971).
  \bibitem{grimvall}G.\ Grimvall, The Electron-Phonon Interaction
    in Metals, (North-Holland, New York, 1981).
  \bibitem{carb1}J.P.\ Carbotte, Rev.\ Mod.\ Phys. {\bf 62},
    1027 (1990).
  \bibitem{daams}J.M.\ Daams and J.P.\ Carbotte, J.\ Low Temp.\
    Phys. {\bf 43}, 263 (1981).
   \bibitem{mcmillan}W.L.~McMillan and J.M.~Rowell, \prl {\bf 19},
    108 (1965).
  \bibitem{mcmillan1}W.L.\ McMillan and J.M.\ Rowell, in
    Superconductivity, ed. R.D. Parks (Marcel Dekker Inc., New York,
    1969), p. 561.
 \bibitem{ru1}A.\ Virosztek and J.\ Ruvalds, \prb {\bf 42},
    4064 (1990).
  \bibitem{ru2}J.\ Ruvalds and A.\ Virosztek, \prb {\bf 43},
    5498 (1991).
  \bibitem{savr}S.Y.\ Savrasov and O.K.\ Andersen, \prl {\bf 77},
    4430 (1996).
  \bibitem{branch}D.\ Branch and J.P.\ Carbotte, \prb {\bf 52},
   603 (1995); J.\ Supercond. {\bf 12}, 667 (1999); Can.\ J.\
   Phys. {\bf 77}, 531 (1999); J.\ Supercond. {\bf 13}, 535
   (2000).
  \bibitem{schach4}J.P.\ Carbotte, E.\ Schachinger, and D.N.\ Basov,
    Nature (London) {\bf 401}, 354 (1999).
  \bibitem{mars4}F.\ Marsiglio, T.\ Startseva, and J.P.\ Carbotte,
    Phys.\ Lett.\ A {\bf 245}, 172 (1998).
  \bibitem{schach7}E.\ Schachinger, J.P.\ Carbotte, and D.N.\ Basov,
    Europhys.\ Lett. {\bf 54}, 380 (2001).
  \bibitem{schach5}E.\ Schachinger and J.P.\ Carbotte, \prb
    {\bf 62}, 9054 (2000).
  \bibitem{schach8}E.\ Schachinger and J.P.\ Carbotte, \prb
    {\bf 64}, 094501 (2001).
  \bibitem{verga}S.~Verga, A.\ Knigavko, and F.~Marsiglio,
    cond-mat/0207145 (unpublished).
  \bibitem{renker}B.~Renker, F.~Gompf, D.~Ewert, P.~Adelmann,
    H.~Schmidt, E.~Gering, and H.~Mutka, Z.\ Phys.\ B {\bf 77},
    65 (1989).
  \bibitem{mars1}F.\ Marsiglio and J.P.\ Carbotte, Aust.\ J.\
    Phys. {\bf 50}, 975 (1997); Aust.\ J.\ Phys. {\bf 50}, 1011
    (1997).
  \bibitem{fong}H.F.\ Fong, P.\ Bourges, Y.\ Sidis, L.P.\ Regnault,
    A.\ Ivanov, G.D.\ Gull, N.\ Koshizuka, and B.\ Keimer,
    Nature (London) {\bf 389}, 588 (1999).
  \bibitem{ros}J.\ Rossat-Mignot, L.P.~Regnault, C.~Vettier,
    P.~Bourges, P.~Burlet, J.~Bossy, J.Y.~Henry, and G.~Lapertot,
    Physica C {\bf 185-189}, 86 (1991).
  \bibitem{mars5}F.~Marsiglio, Molecular Physics Reports {\bf 24},
    73 (1999).
  \bibitem{schach9}E.\ Schachinger and J.P.\ Carbotte, Physica C
    {\bf 364}, 13 (2001).
  \bibitem{dai}P.\ Dai, H.A.\ Mook, S.M.\ Hayden, G.\ Aeppli,
    T.G.\ Perring, R.D.\ Hunt, and F.\ Do\u{g}an, Science {\bf 284},
    1344 (1999).
  \bibitem{he}H.\ He, P.\ Bourges, Y.\ Sidis, C.\ Ulrich, L.P.\ Regnault,
    S.\ Pailh\`es, N.S.\ Berzigiarova, N.N.\ Kolesnikov, and B.\ Keimer,
    Science {\bf 295}, 1045 (2002).
  \bibitem{norm1}M.R.\ Norman and H.\ Ding, \prb {\bf 57}, R11\ 089
  (1998).
  \bibitem{norm2}M.R.\ Norman, M.\ Eschrig, A.\ Kaminski, and J.C.\
  Campuzano, \prb {\bf 64}, 184508 (2001).
  \bibitem{norm3}M.\ Eschrig and M.R.\ Norman, \prb {\bf 67}, 144503 (2003).
\end{thebibliography}
\end{document}